\documentclass[pre,preprintnumbers,amsmath,amssymb,superscriptaddress]{revtex4-1}
\pdfoutput=1

\usepackage{graphicx,color}
\usepackage{bm}
\usepackage{epsfig}

\begin{document}

\title{Spectral content of fractional Brownian motion with stochastic reset}

\author{Satya N. Majumdar}
\address{Laboratoire de Physique Th\'eorique et Mod\`{e}les Statistiques (LPTMS), Universit\'e de Paris-Sud, B\^{a}timent 100, 91405 Orsay Cedex, France}
\author{Gleb Oshanin}
\address{Sorbonne Universit\'e, CNRS, Laboratoire de Physique Th\'{e}orique
de la Mati\`{e}re Condens\'{e}e (UMR 7600), 4 Place Jussieu, 75252 Paris Cedex
05, France}

\begin{abstract}
We analyse the power spectral density (PSD) $S_T(f)$ (with $T$ being the observation time
and $f$ is the frequency) of a fractional Brownian motion (fBm),  
with an arbitrary Hurst index $H \in (0,1)$, undergoing a stochastic resetting to the origin 
at a constant rate $r$
 - the resetting process introduced
some time ago as an example of an efficient, optimisable search algorithm. 
To this end, we first derive an exact expression for the covariance function of 
an arbitrary (not necessarily a fBm) process with a reset, expressing it through the covariance function 
of the parental process without a reset, which yields the desired result for the fBm
in a particular case. We then use this result to compute
exactly the power spectral density for fBM for all frequency $f$.
The asymptotic, large frequency $f$ behaviour of the PSD turns out to be 
distinctly different for sub- $(H < 1/2)$ and super-diffusive $(H > 1/2)$ fBms. 
We show that for large $f$, the PSD has a power law tail: $S_T(f) \sim 1/f^{\gamma}$
where the exponent $\gamma= 2H+1$ for $0<H\le 1/2$ (sub-diffusive fBm), while
$\gamma= 2$ for all $1/2\le H<1$. Thus, somewhat unexpectedly, the exponent $\gamma=2$
in the superdiffusive case $H>1/2$ sticks to its Brownian value and does not depend
on $H$.

\end{abstract}

\maketitle

\section{Introduction}

Power spectral density (PSD) of any stochastic process $X(t)$ provides an important insight into its spectral content and time-correlations \cite{norton}.
The PSD of a real-valued process is standardly defined as 
\begin{equation}
\label{def}
S_T(f) = \frac{1}{T} \int^T_0 dt_2 \int^T_0 dt_1 \cos\left(f \left(t_2 - t_1\right)\right) C(t_1,t_2) \,,
\end{equation}
 where $T$ is the observation time (one usually takes the limit $T \to \infty$),
 $C(t_1,t_2)$ is the auto-correlation function of $X(t)$,
 \begin{equation}
 \label{cov}
C(t_1,t_2) = \langle X(t_1) X(t_2)\rangle \,,
 \end{equation}
and the angle brackets denote the ensemble averaging. 
 As an important property, the PSD was widely studied for various processes across many disciplines, 
 including, e.g., loudness of musical recording \cite{voss,geisel} 
and noise in graphene devices \cite{balandin}, evolution 
of the climate data \cite{talkner} and fluorescence intermittency in nanosystems \cite{fran},  
extremal properties of Brownian motion, such as, the running maximum \cite{oli}, diffusion in an infinite \cite{enzo0} or a periodic Sinai model \cite{enzo},
the time gap between large earthquakes \cite{sornette}  and fluctuations of voltage in nanoscale electrodes \cite{krapf}, 
or of the ionic currents across the nano-pores \cite{ramin1}. 
These are just few stray examples; a more exhaustive list of applications can be found in recent Refs. \cite{eli,we}.

In this paper we analyse the spectral content of the process of fractional
 Brownian motion \cite{ness} with a stochastic reset. 
Its Brownian counterpart has been put forth few years ago in Ref. \cite{EM2011} 
as an example of a robust search algorithm, 
which consists of diffusion tours, 
interrupted  and reset to the origin at random time moments at a fixed rate $r$. 
The mean first passage time of such a process 
to a target, located at some fixed  position in space, was shown to  
non-monotonic function of $r$ and has a deep minimum at some value $r^*$, 
which permits to perform an efficient 
search under optimal conditions. 
Importantly,  a non-zero resetting rate leads to a 
violation of detailed balance, 
and entails a globally current-carrying non-equilibrium 
steady-state with non-Gaussian fluctuations. Different aspects of 
this steady-state and various properties of the diffusion 
process with reset have been extensively 
studied 
\cite{EM2011,1,EM32014,2,EMM2013,MV2013,KMSS2014,GMS2014,MSS2015,MSS22015,Pal2015,3,R2016,NG2016,BEM2016,    
4,5,6,7,HMMT2018}. Very recently, quantum dynamics with resetting to the initial 
state have also been 
studied in various systems~\cite{MSM2018,Garrahan2018}. A different type of resetting 
dynamics (projecting out a measured state from the Hilbert space via successive 
measurements), has been used to compute the first detection probability of a single 
quantum 
particle~\cite{FKB2017,TBK2018}.

Here, we consider a generalisation of the classical version of resetting dynamics, 
for a process
undergoing \textit{fractional} Brownian motion (fBm), as opposed to the
original setting of a standard 
Brownian motion with resetting introduced in Ref. \cite{EM2011}. 
The fBm is a Gaussian process with zero mean and auto-correlation function \cite{ness}
\begin{equation}
\label{fBm}
C_{\rm fBm}(t_1,t_2) = D \left(t_1^{2H} + t_2^{2H} - |t_1 - t_2|^{2H}\right) \,,
\end{equation} 
where $H \in (0,1)$ is the so-called Hurst index and $D$ is the 
proportionality factor with dimension ${\rm length}^2/{\rm time}^{2 H}$. 
For $H > 1/2$ the increments are positively correlated, which results in a 
super-diffusive motion. In contrast, when $H < 1/2$, the increments are 
anti-correlated and one has an anomalous sub-diffusive process. 
The Brownian case is recovered when $H = 1/2$  and 
here $D$ is the usual diffusion coefficient. We let $0<H<1$ to be arbitrary and 
our analysis will cover both cases of super ($1/2\le H<1$)- and sub-diffusion ($0<H\le 1/2$). 
We will show that the asymptotic large frequency behavior of the PSD
$S_T(f)$ has rather different behvaiors in the two cases.

The paper is outlined as follows: In Sec. \ref{auto} we derive a 
general expression 
for the auto-correlation function of an arbitrary process
with stochastic resetting,
expressing it as an integral 
transform of the auto-correlation function of the process without 
resetting.
Next, in Sec. \ref{psd} we focus specifically 
on the frequency-dependence 
of the PSD of a fBm with stochastic reset in the 
limit $T \to \infty$, and also present a general expression 
for its PSD at zero-frequency, valid for an arbitrary observation time $T$.
We conclude with a brief recapitulation of our results and 
an outline of further research in Sec. \ref{disc}.
Details of calculations and $T$-dependent correction terms of the PSD 
are presented in Appendices \ref{A} and \ref{B}.

\section{Auto-correlation function of a reset process} \label{auto} 

In order to compute the PSD defined in Eq. (\ref{def}) for any arbitrary stochastic process, we need to first 
compute the covariance function $C(t_1,t_2)$ of the process in Eq. (\ref{cov}). For a process undergoing 
stochastic reset, the covariance function of the process turns out to be nontrivial. In this section,
we derive an exact expression of the covariance function in presence of reset, in terms of the
covariance function without reset--this relation turns out to be very general and holds for arbitrary 
stochastic process.

\begin{figure} 
\includegraphics[width=0.8\textwidth]{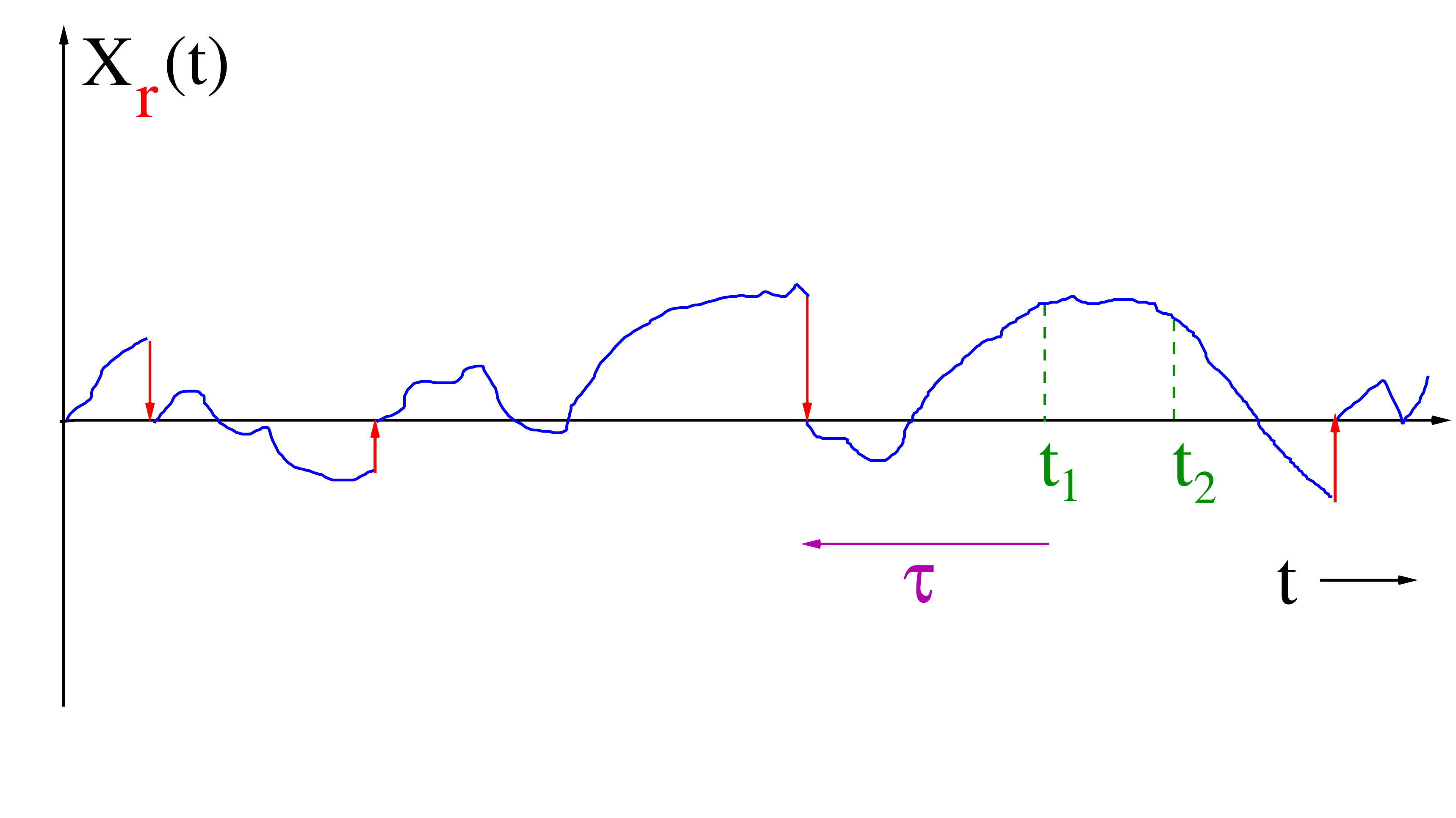} 
\caption{A stochastic process $X_r(t)$ which is 
reset to $0$ with rate $r$. Between two reset events, the process evolves by its own dynamics. Thus resetting 
breaks the process into disjoint `renewal' intervals--following every reset event the process renews itself 
from the origin and evolves by its own natural dynamics till the next reset event. The process is completely 
{\it uncorrelated} from one renewal interval to another. Consider two time epochs $t_1$ and $t_2$, with 
$t_2>t_1$. The correlation function of the reset process $\langle X_r(t_1)X_r(t_2)\rangle$ is identically $0$ 
if the two epochs $t_1$ and $t_2$ belong to two separate renewal intervals. It is nonzero, if and only if, both 
$t_1$ and $t_2$ belong to the same renewal interval.} 
\label{fig.two_point_reset} 
\end{figure}

Let $X_0(t)$ denote an arbitrary stochastic process, starting from $X_0(0)=0$, having zero
mean, $\langle X_0(t)\rangle=0$, and the auto-correlation function $C_0(t_1,t_2)$.  
Now, imagine that the process is interrupted at random times with rate $r$ 
and {\it reset} to $0$. In other words, in a small time interval $dt$, the process is reset to $0$ with 
probability $r\, dt$ and with the complimentary probability $1-r\, dt$ it evolves further by its own natural dynamics. 
Let $X_r(t)$ denote this `reset process' with reset rate parameter $r$. For a typical evolution of the process 
see a sketch in Fig. (\ref{fig.two_point_reset}).

The reset events break the system into disjoint renewal intervals, i.e., 
following every reset event, the process `renews' itself. As a result, 
the correlation of the reset process $X_r(t)$ between two time epochs $t_1$
and $t_2$ (say with $t_1\le t_2$) is identically $0$ if the two epochs
$t_1$ and $t_2$ belong to two separate renewal intervals. The correlation
is nonzero if and only if both $t_1$ and $t_2$ belong to the same renewal
interval (as in Fig. (\ref{fig.two_point_reset})). 
Let us now see how to calculate the auto-correlation 
function of the reset process
$C_r(t_1,t_2)= \langle X_r(t_1)X_r(t_2)\rangle$
in terms of the original auto-correlation function $C_0(t_1,t_2)$ 
(without reset). The crucial observation is 
that for this correlation to be nonzero, $t_1$ and $t_2$ both 
should belong to the same renewal interval, with the convention
$t_1\le t_2$. Let $\tau$ denote the time interval between $t_1$ and
the last reset event that happened before $t_1$ 
(see Fig.  (\ref{fig.two_point_reset})). Then, given that the two epochs 
$t_1$ and $t_2$ belong to the same renewal interval, and given $\tau$,
the correlation function of the reset process would be (considering the fact
that the process restarted at time $\tau$ before $t_1$) just
$C_0(\tau, t_2-t_1+\tau)$. But now the interval $\tau$ itself is a random 
variable drawn from an exponential distribution.
In addition to averaging over all possible $\tau$, we also have to ensure
that there is no reset event between $t_1$ and $t_2$: this happens with 
probability $e^{-r\, (t_2-t_1)}$ for $t_2\geq t_1$. Gathering all the probability events, we then find that
\begin{eqnarray}
C_r(t_1,t_2)= e^{-r\, (t_2-t_1)}\left[ \int_0^{t_1} d\tau\, r\, e^{-r\tau}\, 
C_0(\tau, t_2-t_1+\tau) + e^{-r\,t_1}\, C_0(t_1,t_2)\right]\, .
\label{reset_auto.1}
\end{eqnarray}
This result is easy to understand. The overall factor $e^{-r (t_2-t_1)}$ indicates the probability that there
is no reset event between $t_1$ and $t_2$. The first term inside the square bracket corresponds to the event
that the last reset before $t_1$ occurs between $\tau$ and $\tau+d\tau$ (with probability $r\, e^{-r \tau}\, d\tau$).
The second term corresponds to the event that there was no reset event before $t_1$, in which case the correlation
between $X_r(t_1)$ and $X_r(t_2)$ is the same as $C_0(t_1,t_2)$. 

Particularly simple expression for the auto-correlation function $C_r(t_1,t_2)$ obtains in the case when the original process $X_0(t)$ is an ordinary Brownian motion.
Here, the auto-correlation function is  $C_0(t_1, t_2)= \langle X_0(t_1)X_0(t_2)\rangle= 2D\, {\min}(t_1,t_2)= 2D\, 
t_1$ for $t_1\le t_2$. Hence, Eq. (\ref{reset_auto.1}) gives, for the reset 
process $X_r(t)$, the following auto-correlation function for $t_2\ge t_1$
\begin{equation}
C_r(t_1,t_2)= \frac{2D}{r}\, e^{-r(t_2-t_1)}\, \left[1- e^{-r\, t_1}\right]\, .
\label{BM_reset_auto.1}
\end{equation}
If $t_2\le t_1$, one just has to interchange $t_1$ and $t_2$. 

In the case of interest here, i.e.,  when 
$X_0(t)$ is a fBM with an arbitrary Hurst index $H$, whose 
autocorrelation function is given by Eq. (\ref{fBm}), we have for $t_2\ge t_1$
\begin{eqnarray}
C_r(t_1,t_2) &=& D\, e^{-r\, (t_2-t_1)}\,\Big[ r \int_0^{t_1} d\tau\, e^{-r \tau}\, \left(\tau^{2H}+ (\tau+ t_2-t_1)^{2H}   
- (t_2-t_1)^{2H}\right) \nonumber\\
&+& e^{-r\, t_1} \left(t_1^{2H}+t_2^{2H}-(t_2-t_1)^{2H}\right)\Big]\, .
\label{fbm_reset_auto.1}
\end{eqnarray}
Once again, for $t_2\le t_1$, we need to interchange $t_1$ and $t_2$. Note that for $H\neq 1/2$, the process 
is non-Markovian. However, the result in Eq. (\ref{reset_auto.1}) is very general, and holds even for 
non-Markovian processes (the important point is that every reset event renews the process, and 
$X_r(t)$  does not remember
its history prior to the last resetting epoch before $t$).

\section{The power-spectral density of the reset process}
\label{psd}

We turn next to the analysis of the PSD in Eq. (\ref{def}) with the 
auto-correlation function defined by Eqs. (\ref{BM_reset_auto.1}) and 
(\ref{fbm_reset_auto.1}).
%
It is expedient first to start with the case of an ordinary 
Brownian motion ($H=1/2$).  
Here, the PSD defined by Eqs. (\ref{def}) and (\ref{BM_reset_auto.1}) 
can be straightforwardly calculated
for an arbitrary observation time $T$ to give
\begin{equation}
S_T(f) = \frac{4D}{f^2+r^2}\left[1- \frac{1}{fT(f^2+r^2)}\,
\left(2\,f\,r- 2\,f\,r\, \cos(fT)\, e^{-r\,T}+ (f^2-r^2)\, \sin(fT)\, e^{-r\,T}\right)\right]\, .
\label{psd_bm_reset}
\end{equation}
From this exact expression, one can work out various limiting cases.
For example, keeping $T$ fixed,
we can check that for $r=0$, the expression in Eq. (\ref{psd_bm_reset}) reduces to the well known result for the ordinary Brownian motion
\begin{equation}
S_T(f) \xrightarrow[r=0]{} \frac{4D}{f^2}\, \left[1- \frac{\sin(fT)}{fT}\right]\, .
\label{psd_bm}
\end{equation}
In contrast, keeping $r$ fixed, and taking $T\to \infty$ 
limit, PSD in Eq. (\ref{psd_bm_reset}) reduces to a Lorenzian
as a function of the frequency $f$
\begin{equation}
 S_T(f) \xrightarrow[T\to \infty]{} \frac{4D}{f^2+r^2}  \, ,
\label{lorentz_bm_reset}
\end{equation}
which reflects the fact that at long times 
$X_r(t)$ becomes a stationary process in the
presence of a non-zero reset rate $r$~\cite{EM2011}. 
Interestingly, in this limit the PSD has exactly the 
same Lorenzian form as the PSD of the Ornstein-Uhlenbeck 
process (see, e.g., Ref. \cite{berg}), 
indicating that the stochastic resetting dynamically generates
an effective restoring force. 

One can also make another interesting nontrivial
check. Note that in the limit $f\to 0$,
\begin{equation}
S_T(f=0) = \frac{1}{T}\, \left\langle \left[\int_0^{T} X_r(t)\, dt\right]^2 \right\rangle
= \frac{1}{T} \langle A(T)^2\rangle\, ,
\label{f0_area_bm.1}
\end{equation}
where $A(T)= \int_0^T X_r(t)\, dt$ is just the area under the reset process up to time $T$.
Hence, up to a global prefactor $1/T$,  $S_T(f=0)$ can be thought of as the variance of the area 
under the reset process over the interval $[0,T]$ (note that the mean area $\langle A(T)\rangle=0$).
Hence, from our exact formula in Eq. 
(\ref{psd_bm_reset}), we get by taking $f\to 0$ limit
\begin{equation}
S_T(f=0)= \frac{4D}{r^3 T}\, \left[rT-2+(2+rT)\, e^{-rT}\right]\, .
\label{f0_area_bm.2}  
\end{equation}
This formula matches exactly (after multiplying by $T$) with the formula for the second moment of the area
under a reset process up to time $T$ that was computed recently~\cite{HMMT2018}.

Consider next the PSD $S_T(f)$ of the fBm process with reset
for generic $H\in (0,1)$, whose auto-correlation function is defined in 
Eq. 
(\ref{fbm_reset_auto.1}). A detailed analysis of the full expression for 
the PSD, including all the
$T$-dependent corrections, is presented in Appendix \ref{A}. 
Here we just present the leading term in the large $T$ limit
that reads
\begin{equation}
\label{psdmain}
S_T(f) \xrightarrow[T\to \infty]{}  2\,D\, \Gamma(1+2H)  
\left(\frac{1}{r^{2H-1}} \frac{1}{r^2 + f^2} + \frac{\sin\left(2 H \arctan\left(f/r\right)\right)}{f} \frac{1}{\left(r^2 + f^2\right)^H} \right) \,.
\end{equation}
Interestingly, this expression stems entirely from the first term in Eq. (\ref{fbm_reset_auto.1}), which accounts for the multiple resetting events and hence, is a characteristic feature of the 
resetting process. The contribution associated with the second term in Eq. (\ref{fbm_reset_auto.1}), which is conditioned by the event that no reset 
occurs before $t_1$, and hence, is more specific  to the correlation 
properties of a single tour of a fBm process,  
vanishes in the limit $T \to \infty$. For $H = 1/2$, the expression in Eq. (\ref{psdmain}) coincides with Eq. (\ref{lorentz_bm_reset}) above.

The result in Eq. (\ref{psdmain}) shows that the 
limiting (as $T \to \infty$) form of the PSD for a fBm with an arbitrary 
$H$ 
is a sum of two contributions : (i) a standard Lorenzian, as in
the case of a standard
Brownian motion with reset in Eq. (\ref{lorentz_bm_reset}),  
but now with an $H$-dependent amplitude $\Gamma(1+2H)\, r^{1 - 2H}$, 
which vanishes 
when $r \to 0$ in the sub-diffusive case, and 
diverges in the case of a super-diffusive motion. The latter circumstance 
can be easily understood 
since the PSD of a super-diffusive fBm is time-dependent, and 
diverges as $T^{2H-1}$ in the limit $T \to \infty$ \cite{we2}.  A more detailed discussion of the ageing behaviour of the PSD in terms of a generalized
Wiener-Khinchin theorem for non-stationary processes can be found in Ref. \cite{eli1}.
(ii) the second contribution is 
a Lorenzian in power $H$, 
modulated by the sine term 
divided by the frequency $f$. 
This latter factor converges to $\sin(\pi H)/f$ when $r \to 0$, for any $H$. 

Another interesting limit is the high-frequency regime ($f\to \infty$) at 
a fixed $r$, 
which probes the spectral content of short tours of a fBm with a reset. We observe that in this limit the first contribution vanishes universally as $1/f^2$, while the second one exhibits an $H$-dependent 
decay of the form $1/f^{2 H+1}$. Respectively, this implies that for a sub-diffusive fBm with reset, we find
\begin{equation}
S_T(f) \xrightarrow[T\to \infty, f \to \infty]{}  
\frac{2\,D\, \Gamma(1+2H) \sin(\pi H)}{f^{2H+1}}\, ; \quad\quad H<1/2
\end{equation}
which is independent of the reset rate $r$ and indeed 
coincides 
with the PSD of the sub-diffusive fBm without resetting \cite{we2,mart}.  
On the other hand, for the PSD of a super-diffusive fBm we 
find the following asymptotic form
\begin{equation}
S_T(f) \xrightarrow[T\to \infty, f \to \infty]{}  \frac{2\,D\, 
\Gamma(1+2H) 
}{r^{2H-1}} \frac{1}{f^2}\, ; \quad\quad H>1/2
\end{equation}
with a universal exponent $2$, independent of the actual value of $H$. 
We note that this anomalous frequency-dependence 
has been predicted for a super-diffusive fBm without reset 
(in which case the amplitude is a growing function of $T$)
and also observed experimentally for the dynamics of amoeba and their vacuoles in Ref. \cite{we2}.
It was called 'deceptive' in Ref. \cite{we2},  
since it may lead to an incorrect 
conclusion that one is observing a standard Brownian motion, 
which is certainly not the case. Summarizing, in large frequence $f\to \infty$ limit,
the PSD has a power law tail: $S_T(f)\sim f^{-\gamma}$ where the exponent 
\begin{equation}
\gamma= \begin{cases} 2H+1 \quad \quad\,\, {\rm for}\quad 0< H \le 1/2 \nonumber \\
2 \quad\quad\quad\quad {\rm for} \quad 1/2 \le H <1
\end{cases}
\label{gamma_exponent}
\end{equation}

Lastly, we  
generalise the zero-frequency PSD in Eq. (\ref{f0_area_bm.2}) for a 
fBm with an arbitrary Hurst index $H$. 
Relegating the details of calculations  to Appendix \ref{B}, 
we present below the following exact expression, (which we conveniently order with respect to the behaviour 
of the corresponding terms in the limit $T \to \infty$), 
\begin{eqnarray}
\label{f0_area_fbm.2}  
S_T(f=0) &=& \frac{2\,D\, \Gamma(2 H + 2) }{r^{2 H + 1}} \left(1 - 
\frac{\left(2 H +1\right)}{r T}\right) + \frac{r D T^{2H+2}}{H+1} e^{- r T} \nonumber\\
&-& \frac{D}{\left(H + 1\right) r^{2 H +1}} \left(1 - \frac{2 \left(H + 
1\right)}{r T}\right) \Gamma(2 H + 3, r T) - \frac{2\, D\, \Gamma(2 H + 2, 
r T)}{r^{2 H + 2} T}  \,, \end{eqnarray}
where $\Gamma(a,x)$ is the upper incomplete Gamma-function.  Note that Eq. (\ref{f0_area_fbm.2}) is valid for any $H$, $T$ and $r$. Setting $H = 1/2$, we recover from Eq. (\ref{f0_area_fbm.2}) the result in Eq. (\ref{f0_area_bm.2}). Further on, letting $r \to 0$ at a fixed $T$, we get 
\begin{equation}
S_T(f=0) =  \frac{D T^{2 H + 1}}{H + 1} -  
\frac{(2 H+1) D T^{2H+2}}{2 H^2 + 5 H+3} r + O\left(r^2\right) \,,
\end{equation}
which implies that the variance $\langle A(T)^2\rangle$ of the area $A(T)$
under the  fBm process without resetting ($r=0$ ) grows in proportion to $T^{2 H}$, as it should. For $T \to \infty$ at a fixed $r >0$, we find that $S_T(0)$ in this limit is given by the first term in Eq. (\ref{f0_area_fbm.2}), i.e., 
\begin{equation}
S_T(f=0) = \frac{2\,D\, \Gamma(2 H + 2)}{r^{2 H + 1}} \left(1 - 
\frac{\left(2 H +1\right)}{r T}\right) + O\left(e^{- r T}\right) \,.
\end{equation}
The leading in this equation behaviour is fully compatible with Eq. (\ref{psdmain}).

\section{Conclusion}
\label{disc}

To summarise, we studied here the spectral content of a fractional Brownian motion with an arbitrary Hurst 
index $H$, subject to a stochastic reset at a fixed rate $r$. To this end, we first focused on the 
autocorrelation function $C_r(t_1,t_2)$ of the process with reset and evaluated an exact form of such a 
function, valid for arbitrary values of the parameters characterising our model. As a matter of fact, the 
derived expression has a much broader range of validity (than only a fBm) and holds for an 
\textit{arbitrary} stochastic process with a stochastic reset, expressing its autocorrelations 
$C_r(t_1,t_2)$ through the latter of the unperturbed process, i.e., $C_{r=0}(t_1,t_2)$. 
Using this autocorrelation function, we have computed an exact form of the power spectral 
density $S_T(f)$ of a fBm with stochastic reset in the limit $T \to \infty$. We have shown that the latter 
is a sum of two terms: a standard Lorenzian function and a Lorenzian in power $H$. As a consequence, the large-$f$ 
asymptotic behaviour of $S_{\infty}(f)$ appears to be distinctly different for sub- and super-diffusive 
fBms: For $0<H \le 1/2$, we found that $S_{\infty}(f) \sim 1/f^{2 H +1}$, likewise the parental fBm process, 
with an amplitude independent of the reset rate. Surprisingly, in the super-diffusive case ($H > 1/2$) 
$S_{\infty}(f)$ is described by a universal law $S_{\infty}(f) \sim 1/f^2$, regardless of the actual value 
of $H \geq 1/2$, i.e., has of a form of the spectrum of a standard Brownian motion. In this case, however, 
the amplitude is dependent on the reset rate $r$ and diverges when $r \to 0$.

A natural continuation of our work is to consider a power spectral density of an individual trajectory of a 
fractional Brownian motion with a stochastic reset. Similarly to the analysis presented in Refs. \cite{we} 
and \cite{we2} for Brownian motion and fractional Brownian motion without a reset, we plan to evaluate the 
variance and the full probability density function of such a random variable, parametrised by frequency, 
the observation time and the reset rate.  

\section*{Acknowledgments}

The authors wish to thank the warm hospitality of the SRITP, the Weizmann 
Institute of Science, Rehovot, Israel, where this work was initiated 
during the workshop ``Correlations, fluctuations and anomalous transport 
in systems far from equilibrium" held in December, 2017.

\appendix
\section{Details of the derivation of the result in Eq. (\ref{psdmain}).}
\label{A}

Consider the contribution to the PSD of the fBm process 
with reset, which stems out of the 
first term in Eq. (\ref{fbm_reset_auto.1}). This 
contribution  is given explicitly by
\begin{equation}
S_1 = \frac{2 r D}{T} \int^T_0 dt_2 \int^{t_2}_0 dt_1 \cos\left(f \left(t_2 - t_1\right)\right) e^{- r (t_2 - t_1)} \int^{t_1}_0 d\tau 
e^{-r \tau} \left[\tau^{2 H} + (\tau + t_2 - t_1)^{2H} - \left(t_2 - t_1\right)^{2H}\right] \,.
\end{equation}
Changing the integration variables $\tau \to t_1 \phi$ and then, $t_1 \to t_2 \xi$, we rewrite the latter expression as
\begin{equation}
S_1  = \frac{2 r D}{T} \int^T_0 t_2^{2H+2} dt_2 \int^{1}_0 \xi d\xi \cos\left(f t_2 \left(1 - \xi\right)\right) \, \int^{1}_0 d\phi \, e^{- r t_2 (1 - \xi + \xi \phi)}  
 \left[\xi^{2 H} \phi^{2H} + (1 + \xi \phi - \xi)^{2H} - \left(1 - \xi\right)^{2H}\right] \,.
\end{equation}
At the next step, we expand both the exponential and the cosine terms in the Taylor series in powers of $f$ and $r$, respectively, 
and integrate over $t_2$ 
to get
\begin{equation}
\label{z}
S_1  = 2 r D  T^{2H + 2} \sum_{n=0}^{\infty} \frac{(-1)^n}{(2 n)!} \left( f T\right)^{2 n} \sum_{m=0}^{\infty} \frac{(-1)^m}{m!} \left(r T\right)^m \frac{A_{n,m}}{\left(3 + 2H + m + 2n\right)} \,,
\end{equation}
with
\begin{eqnarray}
A_{n,m} &=& \int^1_0 \xi d\xi \left(1 - \xi\right)^{2 n} \int^1_0 d\phi \left(1 - \xi + \xi \phi\right)^m \left[ \xi^{2H} \phi^{2H} + \left(1-\xi+\xi \phi\right)^{2H} - \left(1 - \xi\right)^{2H}\right]= \nonumber\\
&=& \frac{\Gamma(1+2H) (2n)!}{\left(2+2H+m+2n\right) \Gamma(2+2H+2n)} + \frac{1}{(1+2n) \left(2+2H+m+2n\right)} - \nonumber\\
&-& \frac{1}{\left(1 + 2 H + 2n\right) \left(2+2H+m+2n\right)} \,.
\end{eqnarray}
Now, we can straightforwardly perform summation over $m$, which gives
\begin{eqnarray}
\label{z1}
S_1  &=& r D T^{2H + 2} \sum_{n=0}^{\infty} \frac{(-1)^n}{(2 n)!} \frac{\left( f T\right)^{2 n}}{(1+H+n)}  \left[\frac{\Gamma(1+2H) (2n)!}{\Gamma(2+2H+2n)} + \frac{1}{(1+2n)} - \frac{1}{\left(1 + 2 H + 2n\right)} 
\right] \nonumber\\
&\times& \left( e^{- r T} + \frac{\Big(r T - 2 (1+H+n)\Big)}{\left(r T\right)^{3+2H+2n}} \Big(\Gamma(3+2H+2n) - \Gamma(3+2H+2n,r T)\Big)
\right)  \,.
\end{eqnarray}

There are several terms in the brackets in the second line in Eq. (\ref{z1}) and we examine their contributions to $S_1$ separately. 
Inspecting each term, we realise that 
the leading large-$T$ behaviour is given by 
\begin{eqnarray}
\label{S1}
&&\frac{2 D}{r^{2H+1}}  \sum_{n=0}^{\infty} \frac{(-1)^n \Gamma(2+2H+2n)}{(2 n)!}  \left(\frac{f}{r}\right)^{2n}  \left[\frac{\Gamma(1+2H) (2n)!}{\Gamma(2+2H+2n)} + \frac{1}{(1+2n)} - \frac{1}{\left(1 + 2 H + 2n\right)} 
\right] =\nonumber\\
&=& 2\,D\, \Gamma(1+2H) \left(\frac{1}{r^{2H-1}} 
\frac{1}{r^2 + f^2} + 
\frac{\sin\left(2 H \arctan\left(f/r\right)\right)}{f} 
\frac{1}{\left(r^2 + f^2\right)^H} \right) \,.
\end{eqnarray} 
Further on, we get
\begin{eqnarray}
&&\frac{2 D}{r^{2H+2} T}  \sum_{n=0}^{\infty} \frac{(-1)^n \Gamma(3+2H+2n)}{(2 n)!}  \left(\frac{f}{r}\right)^{2n}  \left[\frac{\Gamma(1+2H) (2n)!}{\Gamma(2+2H+2n)} + \frac{1}{(1+2n)} - \frac{1}{\left(1 + 2 H + 2n\right)} 
\right] =\nonumber\\
&=& \frac{4 \Gamma(1+2H) D}{r T} \left(\frac{H}{r^{2H-1}} \frac{1}{r^2 + f^2} + \frac{r^{3-2H}}{\left(r^2+f^2\right)^2} + \frac{2 H (1+H)}{\left(r^2+f^2 \right)^{H+1/2}} \,_2F_1\left(H+\frac{1}{2}, - H - \frac{1}{2}, \frac{3}{2}; \frac{f^2}{r^2+f^2}\right)
\right) \,,
\end{eqnarray}
where $_2F_1(\ldots)$ is the Gauss hypergeometric function. This contribution vanishes as $T \to \infty$ and thus defines the leading $T$-dependent corrections to the result in Eq. (\ref{S1}).

Lastly, we notice that the sum 
\begin{eqnarray}
\sum_{n=0}^{\infty} \frac{(-1)^n}{(2 n)!} \frac{\left( f T\right)^{2 n}}{(1+H+n)}  \left[\frac{\Gamma(1+2H) (2n)!}{\Gamma(2+2H+2n)} + \frac{1}{(1+2n)} - \frac{1}{\left(1 + 2 H + 2n\right)} 
\right]
\end{eqnarray}
is bounded from above for any $T$, and hence, the contribution of the first term in the second line in Eq. (\ref{z1}), which contains a factor $\exp(- r T)$, is exponentially small for large $T$ and $r > 0$. In a similar fashion, it is rather straightforward to show that
the terms which contain the upper incomplete gamma-function are also exponentially small when $T \to \infty$ and $r > 0$.

Thus putting everything together, we find that in the large $T$ limit, the 
leading order behaviour of $S_1$ is given by the term in Eq. (\ref{S1}), 
i.e.,
\begin{eqnarray}
S_1\approx 
2\,D\, \Gamma(1+2H) \left(\frac{1}{r^{2H-1}}
\frac{1}{r^2 + f^2} +
\frac{\sin\left(2 H \arctan\left(f/r\right)\right)}{f}
\frac{1}{\left(r^2 + f^2\right)^H} \right) \,.
\label{S1_final}
\end{eqnarray}

Consider next the contribution to the PSD of the fBm process with reset stemming out of the second term in Eq. (\ref{reset_auto.1}). This contribution  is given explicitly by
\begin{equation}
S_2  = \frac{2 D}{T} \int^T_0 dt_2 \, e^{- r t_2} \int^{t_2}_0 dt_1 \, \cos\left(f \left(t_2 - t_1\right)\right) \left(t_1^{2 H} + t_2^{2 H} - \left(t_2 - t_1\right)^{2 H}\right) \,.
\end{equation}
Changing the integration variable $t_1 \to t_2 \zeta$, expanding both the cosine and the exponential terms in Taylor series in the powers of $t_2$ and integrating over this variable, we get
\begin{equation}
\label{B2}
S_2= 2 D T^{2 H + 1} \sum_{n=0}^{\infty} \frac{(-1)^n}{(2 n)!} \left(f T\right)^{2n} \sum_{m=0}^{\infty} \frac{(-1)^m}{m!}  \left(r T\right)^{m}  \frac{B_n}{(2+2H+ 2n+m)} \,,
\end{equation}  
where
\begin{eqnarray}
B_n &=& \int^1_0 d\xi (1-\xi)^{2n} \left(\xi^{2 H} + 1 - (1 - \xi)^{2H}\right) \nonumber\\
&=& \frac{2 H}{(2n+1)(2n+1+2H)}+ \frac{\Gamma(1+2H) (2 n)!}{\Gamma(2 + 2H + 2n)} \,.
\end{eqnarray}
noticing next that the leading 
$T \to \infty$ behaviour of the integral over $t_2$ 
is obtained by extending the upper terminal of integration $T$ to 
infinity, we arrive at the following expression 
\begin{eqnarray}
S_2  =  \frac{2\,D\, \Gamma(2 H + 1)}{r^{2 H} T}  \left(\frac{1}{r^2 + 
f^2} + \frac{r^{2 H-1} \sin\left(2 H \arcsin\left(f/\sqrt{r^2 + f^2}\right)\right)}{f \left(r^2 + f^2\right)^H} \right) + O\left(e^{- r T}\right) \,.
\end{eqnarray}
Note that this contribution vanishes when the observation time $T $ is set equal to infinity.
Hence, $S=S_1+S_2$ with $S_2\to 0$ as $T \to \infty$. In consequence, the leading 
order behaviour for large $T$ is, $S\approx S_1$ with $S_1$ given in Eq. 
(\ref{S1_final}). This completes the derivation of the result
in Eq. (\ref{psdmain}).

\section{Details of the derivation of the result in Eq. (\ref{f0_area_fbm.2}).}
\label{B}

Here we briefly outline the derivation of our result in Eq. (\ref{f0_area_fbm.2}). The contribution to the zero-frequency PSD stemming out of the first term in Eq. (\ref{fbm_reset_auto.1}) can be straightforwardly obtained from Eq. (\ref{z1}) above 
by simply noticing that $S_1(f=0)$ is given by the $n=0$ term in the series. This yields 
\begin{eqnarray}
\label{z7}
S_1(f=0)  &=& \frac{r D T^{2H + 2}}{1+H} \left( e^{- r T} + \frac{\Big(r T - 2 (1+H)\Big)}{\left(r T\right)^{3+2H}} \Big(\Gamma(3+2H) - \Gamma(3+2H,r T)\Big)
\right)  \,.
\end{eqnarray}
Further on, for the contribution stemming out of the second term in Eq. (\ref{fbm_reset_auto.1}) we have from Eq. (\ref{B2})
\begin{eqnarray}
\label{z8}
S_2(f=0) = 2 D T^{2 H + 1} \sum_{m=0}^{\infty} \frac{(-1)^m}{m!}    \frac{\left(r T\right)^{m}}{(2+2H+m)} = \frac{2 D}{r^{2H+2} T} \left(\Gamma(2H+2) - \Gamma(2H+2, r T)\right) \,.
\end{eqnarray}  
Combining the expressions in eqs. (\ref{z7}) and (\ref{z8}) and re-arranging them according to the rate at which they vanish in the limit $T \to \infty$, we get our expression in eq. (\ref{f0_area_fbm.2}). 

\end{document}